\documentstyle[12pt]{article}

\begin{document}

\begin{center}
\LARGE{\bf A Lighthouse Falls} 
\end{center}
\medskip

\begin{center}
Naresh Dadhich \\
Inter-University Centre for Astronomy and Astrophysics, \\
Post Bag 4, Ganeshkhind, Pune 411 007, INDIA
\end{center}

\begin{abstract}
Professor A. K. Raychaudhuri suddenly passed away on 18th June 2005. I have 
attempted to give a glimpse of his life and work, and more specifically his 
role as a great teacher and inspiring force to three generations of students 
of the Presidency College, Kolkata.
\end{abstract}
 
\bigskip

It was hard to take what the computer screen was showing, Amalbabu is no 
more, on that fateful Saturday afternoon on 18th June 2005. It was more so 
for me because I had returned from Kolkata that morning and had seen him on 
the previous Monday, the 13th June. I am referring to the legend of Amal 
Kumar Raychaudhuri, whom we all fondly call AKR and revere him with 
great awe and affection. 

The first message was from Somya Chakraborty which had come via California 
and then a spurt of others followed confirming the inevitable. 
I am referring to the legend of Amal Kumar Raychaudhuri, whom we all 
fondly call AKR and revere him with great awe and affection. In such 
situations, a casual and innocent comment suddenly attains great meaning and 
revelation. This time, he told me not to discuss any physics for he 
said that he made so trivial mistakes in calculations that he could no longer 
trust himself. Mind you, he was a perfectionist and would never make any 
error in calculations as all his students would swear. This innocent statement 
now makes me to reflect, was he telling that his innings was over, no reason 
to go on if he coudn't do physics in the manner he liked to do ?
 
IUCAA and Vigyan Prasar have jointly made a documentary film on him 
wherein he spoke his mind in his characteristic clarity and honesty. 
Fortunately all the shooting of the film was done a couple of weeks 
before the final event. This and the one on another legend, Professor 
P. C. Vaidya will be released in a function at IUCAA on 16th Nov. 2005. 
We hope that the film would be able to transmit to the younger generation 
some bit of his spirit of scholarship and dedication.  

As a teacher, not only you do your work consciously, which is a bare 
minimum, but set a benchmark for performance and a role model for students 
to emulate. AKR was an apt personification of this ideal, and hence 
without question a legend. A good measure of a teacher is reflected in the 
heights scaled by his students. Several of his students are the world class 
and well recognised scientists. As a researcher, he made the 
indelible contribution by his equation - the Raychaudhuri equation governing 
the gravitational dynamics of a bunch of bodies including the whole 
Universe. This will stand firm so long as Einstein's 
theory of gravitation - general relativity (GR) stands. I must say here 
that after the monumental contributions of Jagdish Chandra Bose, C V 
Raman, S N Bose and Meghnad Saha, very few of the Indian contributions make 
this grade.  

AKR was born in Barisal, now in Bangladesh, on 14 September 1923. His 
father was an M Sc with I class in Mathematics and taught in a school in 
Kolkata. Right in school, he demonstrated his aptitude for Mathematics 
when he showed that certain exercise he could do in a simpler way than 
what the teacher had done. This was greatly appreciated by the Headmaster who 
made a mention of it in the school magazine. After school, he joined the 
Presidency College which was later to become his playground for his glorious 
academic innings. He was interested in Maths and Physics with a slight tilt 
towards the former, yet his father citing his own example advised him to do 
the later. He did carry his mathematical bent and flavour all through his 
work by always attempting to prove theorems. The way he derived his equation 
is a brilliant example of it. 

He did his B.Sc. from the Presidency College in 1942 and M.Sc in 1944 from 
Calcutta University, and then started his arduous research journey. He 
joined Indian Association of Cultivation of Science (IACS) in 1945 as a 
research scholar and was asked to do experimental work which he was 
not cut out for. This lasted for four years of sheer frustration and 
agony. There couldn't have been a worse start for a young man's research 
career. Undeterred he taught himself in complete isolation the abstract and 
difficult differential geometry and general relativity (GR). It is this drill 
of self learning which formed the foundation for his later work. 

Then he taught for a couple of years at the Asutosh College. The 
only person who was interested in GR in Kolkata at that time was N R Sen 
but the two did not click as their interests did not match. Once again AKR 
joins IACS in 1952 as a Research Associate with a clear instruction to 
work on the problems assigned by the Director and the head 
of the section. This was indeed the most trying time of his research 
career. GR was not considered worthy of attention of a young researcher 
and he was asked to work on electronic energy bands in metals. He had to 
write two papers just to keep his job at the institute. The path breaking 
equation was discovered in 1954 under such adverse and challenging 
circumstances. Any lesser being could have buckled down under such 
hostility and would have happily carried on studying metals and their 
properties but not AKR. This unflinching strength, courage and commitment to 
his work is what makes AKR stand out not only for his wonderful equation 
but for his sheer audacity and courage - truly a legend.

The liberating event occurs in 1961 when he joins the Presidency College 
as a Professor of theoretical Physics. Then a glorious and most rewarding 
academic career unfolds for nearly three decades. In our enthusiasm to 
advance in scientific research and to catch with the western world, a focused 
and concerted effort was initiated by independent India by establishing 
autonomous research institutes solely devoted to research outside 
the university system. This had as we now realise a very disastrous effect on 
the academic life of universities. For all good and talented scientists were 
drawn to the institutes because of the better lab and infrastructural 
facilities as well as congenial research environment. It has given rise to 
a very anomalous situation, where there are pools of young talented students 
there are no good scientists to teach and train them, and on the other hand 
where there are good scientists there are no students.    

Against this backdrop, people like AKR who are few and far between served 
as lighthouse inspiring and exciting young minds. 

Before discovering his beautiful equation, he wrote one of the very first 
papers on condensations in expanding universe giving rise to 
structures. Also discussed in another paper the question of the 
Schwarzschild singularity which was at that time generally believed, 
including Einstein himself, not to be attainable. He constructed a non static 
collapsing solution and showed that there was nothing to prevent such a 
happening and thereby 
challenging the prevelant view. He was certainly right as we now know that 
there was nothing singular about the Schwarzschild singularity. It was 
simply an artifact of a bad coordinate choice. It however defined the 
black hole horizon, which had interesting physical property that nothing 
could come out of it. These were also very important and interesting 
papers which were done while he was working as a temprorary lecturer in 
Asutosh College, perhaps a feat unparalleled in the history of that college or 
any college.

The Schwarzschild radius was non singular but there was a genuine physical 
singularity as the radius goes to zero where spacetime curvature diverges. 
The Oppenheimer - Snyder solution of collapsing homogeneous dust ball, and 
also his own solution of collapsing sphere, ultimately lead to this singular 
state at the centre with density and curvatures diverging. In the cosmology 
too there was the big-bang singularity of the 
homogeneous isotropic and expanding Friedman - Robertson - Walker 
model. As the universe is now expanding which means it should have 
been very compact in the past. Judging by the present rate of expansion it 
is clear that it should have had a singular and explosive beginning in a 
big-bang about 13 billion years ago. 

The natural question to ask was whether singularity of gravitational collapse 
or of expanding universe is due to homogeneity and isotropy or generic and 
inherent character of GR, Einstein's theory of gravitation ? For instance, 
could rotation which opposes gravity avert its occurrence. AKR was set 
on this track by the Goedel's static rotating model. He then 
addressed the fundamental question of singularity in the most general form 
with no reference to any symmetry and any specific property of spacetime 
and energy distribution. He considered evolution of a congruence of 
ordinary particles, which are characterised by timelike velocity vector,  
under its own gravity. Taking the timelike velocity as the eigenvector 
of the Ricci curvature and then by using the Einstein equation, he obtained his 
celebrated equation  - the Raychaudhuri Equation. It related the rate of 
expansion of a congruence of freely moving particles with the expansion 
(divergence), shear (distortion) and rotation as well as with 
the energy density (including pressure) which pulls particles together. 
Here shear works hand in hand with the energy density in pulling things 
together while rotation opposes it. The most profound result emanating from 
the equation was that singularity is a generic and inevitable feature of 
GR. It was left to the mathematical prowess of Roger Penrose and Stephen 
Hawking to prove in mid sixties the most general and powerful 
theorems to establish this result rigorously.  The Raychaudhuri equation 
was the key to the most powerful and general result: under certain 
reasonable assumptions on energy 
conditions and causality otherwise in all generality, singularity is 
unavoidable in GR. 

In 1990, J M M Senovilla shocked the relativity community by obtaining a 
singularity free cosmological solution of the Einstein equation. How did 
it happen, how did it bypass the all encompassing singularity theorems ? 
The assumptions of the theorems looked quite obvious and natural except 
for the one on existence of trapped surfaces (surface from which even 
light can't come out), which was always a suspect. It was particularly 
invoked in the context of collapse to black hole and was quite alright 
for that situation. Senovilla's solution violated this assumption and 
hence the theorems became ineffective but this violation entailed no 
unphysical feature. His early reaction to singularity free models was of 
caution and hesitation but over the period he came around and I coauthored a 
paper with him on a spherical cosmological model (Senovilla's was 
cylindrical) without big-bang singularity. He again addressed the question in 
general terms and obtained the necessary condition for singularity free 
models as the vanishing of space and time averages of all physical and 
geometrical parameters.  
This result was published in the Physical Review Letters. His 
last paper dealt with the singularity free perfect fluid models, and he 
was working to include rotation.

It is only in late fifties when he learnt that his paper was  
much talked of in the west and was referred by Jordan and Heckmann that he 
gained confidence to submit his thesis for D. Sc. in 1959. The great GR 
guru, John Wheeler was one of the examiners and recommended that if there 
was a provision for a degree with Honours, then here was the one. First 
recognition came from far, and yet it did not make much news at home until 
Jayant Narlikar's return to India in 1972. It is only then AKR surfaced on the 
Indian scene and slowly Academies and other academic agencies started 
taking note of him. 

It is noteworthy that though he was an icon for his students yet not many 
of his good students took to research in GR. This was perhaps because he 
thought that there was greater excitement and action in other fields like 
high energy and condensed matter physics. He was indeed a very honest and 
true scholar of highest intellect and integrity. Scholarship was his sole 
driving force through the hard times. Shobo Bhattacharya, TIFR Director 
told us that AKR was only three sentences away for his students scattered 
all over the globe. When one returned from a 
visit to Kolkata, the enquery would proceed in this order, how did you 
find Kolkata, did you go to the Coffee House and how was AKR ?

AKR symbolised the spirit of scholarship, excellent teaching and 
research in college/university, that shined like a lighthouse. A true 
and heartful tribute to him would be not to let this spirit wane and fade. 
Finally, had he met with encouragement and appreciation in his early 
research career, things would have certainly been different. Least of all 
Calcutta University won't have remained oblivious of him. In fifties, he 
would have been in scientific currency internationally and on the top of the 
recognition ladder. More importantly if he could have had the benefit of a 
good mathematical group equipped with the sophisticated techniques 
of differential geometry and global analysis, it was quite possible that 
he could have before the advent of Penrose and Hawking on the 
scene, proved the famous singularity theorems. That would have been 
really remarkable and could perhaps have changed the course of 
gravitational and theoretical physics research in the country. But then we 
won't have had the legend of AKR.  

\end{document}